\documentclass[reprint, superscriptaddress, twocolumn, amsmath, amssymb, aps, prb]{revtex4-2}
\usepackage{graphicx}
\usepackage{dcolumn}
\usepackage{bm}
\usepackage{placeins}
\usepackage{appendix}
\usepackage{upgreek}

\usepackage{verbatim}
\usepackage[usenames,dvipsnames]{xcolor}
 \usepackage{amsmath}
 \usepackage{amsfonts}
 \usepackage{amssymb}

 \usepackage{bbold}     
 \usepackage[makeroom]{cancel}  
 \usepackage{multirow}    
 \usepackage[normalem]{ulem}        
 \usepackage{array}
 \usepackage{pdfpages}
\makeatletter
 \AtBeginDocument{\let\LS@rot\@undefined}
 \makeatother

\begin{document}


\title{Selective-Area-Grown PbTe-Pb Planar Josephson Junctions for Quantum Devices}

\author{Ruidong Li}
\email{R.L., W.S., W.M., and Z.Y. contributed equally to this paper.}
\affiliation{State Key Laboratory of Low Dimensional Quantum Physics, Department of Physics, Tsinghua University, Beijing 100084, China}

\author{Wenyu Song}
\email{R.L., W.S., W.M., and Z.Y. contributed equally to this paper.}
\affiliation{State Key Laboratory of Low Dimensional Quantum Physics, Department of Physics, Tsinghua University, Beijing 100084, China}

\author{Wentao Miao}
\email{R.L., W.S., W.M., and Z.Y. contributed equally to this paper.}
\affiliation{State Key Laboratory of Low Dimensional Quantum Physics, Department of Physics, Tsinghua University, Beijing 100084, China}

\author{Zehao Yu}
\email{R.L., W.S., W.M., and Z.Y. contributed equally to this paper.}
\affiliation{State Key Laboratory of Low Dimensional Quantum Physics, Department of Physics, Tsinghua University, Beijing 100084, China}

\author{Zhaoyu Wang}
\affiliation{State Key Laboratory of Low Dimensional Quantum Physics, Department of Physics, Tsinghua University, Beijing 100084, China}

\author{Shuai Yang}
\affiliation{State Key Laboratory of Low Dimensional Quantum Physics, Department of Physics, Tsinghua University, Beijing 100084, China}

\author{Yichun Gao}
\affiliation{State Key Laboratory of Low Dimensional Quantum Physics, Department of Physics, Tsinghua University, Beijing 100084, China}

\author{Yuhao Wang}
\affiliation{State Key Laboratory of Low Dimensional Quantum Physics, Department of Physics, Tsinghua University, Beijing 100084, China}

\author{Fangting Chen}
\affiliation{State Key Laboratory of Low Dimensional Quantum Physics, Department of Physics, Tsinghua University, Beijing 100084, China}

\author{Zuhan Geng}
\affiliation{State Key Laboratory of Low Dimensional Quantum Physics, Department of Physics, Tsinghua University, Beijing 100084, China}

\author{Lining Yang}
\affiliation{State Key Laboratory of Low Dimensional Quantum Physics, Department of Physics, Tsinghua University, Beijing 100084, China}

\author{Jiaye Xu}
\affiliation{State Key Laboratory of Low Dimensional Quantum Physics, Department of Physics, Tsinghua University, Beijing 100084, China}

\author{Xiao Feng}
\affiliation{State Key Laboratory of Low Dimensional Quantum Physics, Department of Physics, Tsinghua University, Beijing 100084, China}
\affiliation{Beijing Academy of Quantum Information Sciences, Beijing 100193, China}
\affiliation{Frontier Science Center for Quantum Information, Beijing 100084, China}
\affiliation{Hefei National Laboratory, Hefei 230088, China}

\author{Tiantian Wang}
\affiliation{Beijing Academy of Quantum Information Sciences, Beijing 100193, China}
\affiliation{Hefei National Laboratory, Hefei 230088, China}

\author{Yunyi Zang}
\affiliation{Beijing Academy of Quantum Information Sciences, Beijing 100193, China}
\affiliation{Hefei National Laboratory, Hefei 230088, China}

\author{Lin Li}
\affiliation{Beijing Academy of Quantum Information Sciences, Beijing 100193, China}

\author{Runan Shang}
\affiliation{Beijing Academy of Quantum Information Sciences, Beijing 100193, China}
\affiliation{Hefei National Laboratory, Hefei 230088, China}

\author{Qi-Kun Xue}
\affiliation{State Key Laboratory of Low Dimensional Quantum Physics, Department of Physics, Tsinghua University, Beijing 100084, China}
\affiliation{Beijing Academy of Quantum Information Sciences, Beijing 100193, China}
\affiliation{Frontier Science Center for Quantum Information, Beijing 100084, China}
\affiliation{Hefei National Laboratory, Hefei 230088, China}
\affiliation{Southern University of Science and Technology, Shenzhen 518055, China}

\author{Ke He}
\email{kehe@tsinghua.edu.cn}
\affiliation{State Key Laboratory of Low Dimensional Quantum Physics, Department of Physics, Tsinghua University, Beijing 100084, China}
\affiliation{Beijing Academy of Quantum Information Sciences, Beijing 100193, China}
\affiliation{Frontier Science Center for Quantum Information, Beijing 100084, China}
\affiliation{Hefei National Laboratory, Hefei 230088, China}

\author{Hao Zhang}
\email{hzquantum@mail.tsinghua.edu.cn}
\affiliation{State Key Laboratory of Low Dimensional Quantum Physics, Department of Physics, Tsinghua University, Beijing 100084, China}
\affiliation{Beijing Academy of Quantum Information Sciences, Beijing 100193, China}
\affiliation{Frontier Science Center for Quantum Information, Beijing 100084, China}


\begin{abstract}

\centerline{\bf Abstract \rm}

Planar Josephson junctions are predicted to host Majorana zero modes. The material platforms in previous studies are two dimensional electron gases (InAs, InSb, InAsSb and HgTe) coupled to a superconductor such as Al or Nb. Here, we introduce a new material platform for planar JJs, the PbTe-Pb hybrid. The semiconductor, PbTe, was grown as a thin film via selective area epitaxy. The Josephson junction was defined by a shadow wall during the deposition of the superconductor Pb. Scanning transmission electron microscopy reveals a sharp semiconductor-superconductor interface. Gate-tunable supercurrent and multiple Andreev reflections are observed.  A perpendicular magnetic field causes interference patterns of the switching current, exhibiting Fraunhofer-like and SQUID-like behaviors. We further demonstrate a prototype device for Majorana detection, wherein phase bias and tunneling spectroscopy are applicable.

{\bf KEYWORDS: \emph{PbTe, planar Josephson junctions, supercurrent interference, topological superconductivity}\rm}

\end{abstract}

\maketitle  

Semiconductor-superconductor hybrids with strong spin-orbit interaction can exhibit intriguing phases of matter, notably topological superconductivity \cite{ReadGreen, Kitaev}. One promising platform involves coupling a two-dimensional electron gas (2DEG) with a superconductor to create a planar Josephson junction (JJ). Majorana zero modes are predicted to localize near the two edges of the JJ  \cite{PJJ_PRL_2017, PJJ_PRX_2017}. In comparison to nanowire platforms \cite{Lutchyn2010, Oreg2010, WangZhaoyu}, the key advantage of this platform is the phase bias across the junction. This additional experimental knob can reduce the critical magnetic field required for Majoranas. Previous studies have employed HgTe, InAs, InSb and InAsSb as the 2DEG materials \cite{HgTe, 2016_PRB_Palmstrom, InSb_2DEG_2015, InAsSb_2DEG_2021}. Appealing experimental progress, including supercurrent interference \cite{Yacoby_Fraunhofer, Vlad_NatureNano, 2017_PRB_Fraunhofer,Shabani_NC,  2021_Shabani_InAs_Fraunhofer}, $\pi$-junction physics \cite{Yacoby_pi, Srijit_pi, Shabani_pi} and zero-bias conductance peaks \cite{2019_Nature_Charlie, 2019_Nature_Yacoby, 2023_PRL_Charlie_PJJ, 2023_PRB_Charlie_PJJ} have been demonstrated in corresponding devices. Theoretical studies have systematically explored the physics of planar JJs and the roadmap toward braiding and topological quantum computation \cite{LiuJie_PJJ, Wimmer_PJJ, Tudor_PJJ, Zhou_PJJ}. The absence of end-to-end correlations of zero-bias peaks implies that the current experiments are limited by disorder \cite{2023_PRB_Charlie_PJJ}. Eliminating disorder is of critical importance for Majorana experiments and topological quantum computation \cite{Patrick_Lee_disorder_2012, GoodBadUgly}. 

Here, we present the PbTe-Pb hybrid as a new planar JJ material platform for exploring topological superconductivity and other potential applications, such as gate tunable qubits \cite{2015_PRL_gatemon, DiCarlo_gatemon,Huo_gatemon}. While the PbTe-Pb hybrid has recently been proposed in its nanowire form as a Majorana candidate \cite{CaoZhanPbTe} with promising  experimental progress \cite{Jiangyuying, Erik_PbTe_SAG, PbTe_AB, Fabrizio_PbTe, Zitong, Wenyu, Yichun, Yuhao}, in this paper, we realize its 2DEG version, motivated by the Majorana proposals for topological planar JJs \cite{PJJ_PRL_2017, PJJ_PRX_2017}. Selective area growth (SAG) of PbTe film has been previously achieved on InP, a substrate that is lattice-mismatched with PbTe \cite{Erik_PbTe_film}. In this work, the SAG PbTe film is grown on a lattice-matched substrate to minimize disorder. The junction was defined using a shadow wall during the in-situ growth of the superconductor. This process prevents direct etching of the superconductor on the junction. Subsequent fabrication steps also kept the junction region intact, further mitigating  sources of disorder. Transport characterization of the planar JJs reveals a gate-tunable supercurrent and its interference in a perpendicular magnetic field. In the end, we demonstrate the design of a prototype device for Majorana detection.

\begin{figure*}[htb]
\includegraphics[width=\textwidth]{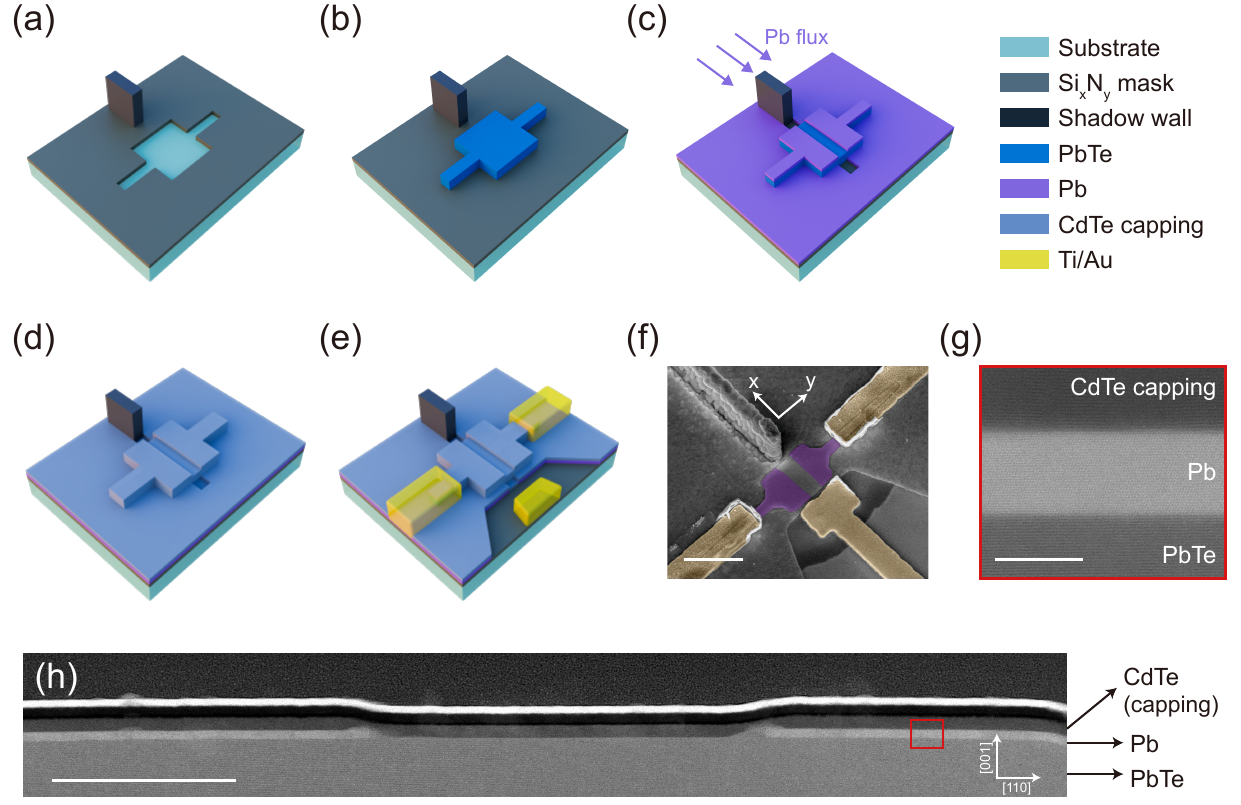}
\centering
\caption{Selective area growth of PbTe-Pb planar JJs. (a) The CdTe/Pb$_{1-x}$Eu$_x$Te substrate is covered by a SiN dielectric mask (dark grey) with an HSQ shadow wall. Part of the mask was etched for the following selective growth. (b) SAG of the PbTe film. (c) In situ deposition of the Pb superconductor. The shadowed region defines the JJ. (d) Growth of the CdTe capping layer. (e) Schematic of a final device with contacts and side gates. (f) Tilted SEM of a typical device. The scale bar is 1 $\upmu$m. (g) STEM of the interfaces. The scale bar is 10 nm. (h) STEM of the JJ cross-section. The scale bar is 200 nm.  }
\label{fig1}
\end{figure*}

\begin{figure}[htb]
\includegraphics[width=\columnwidth]{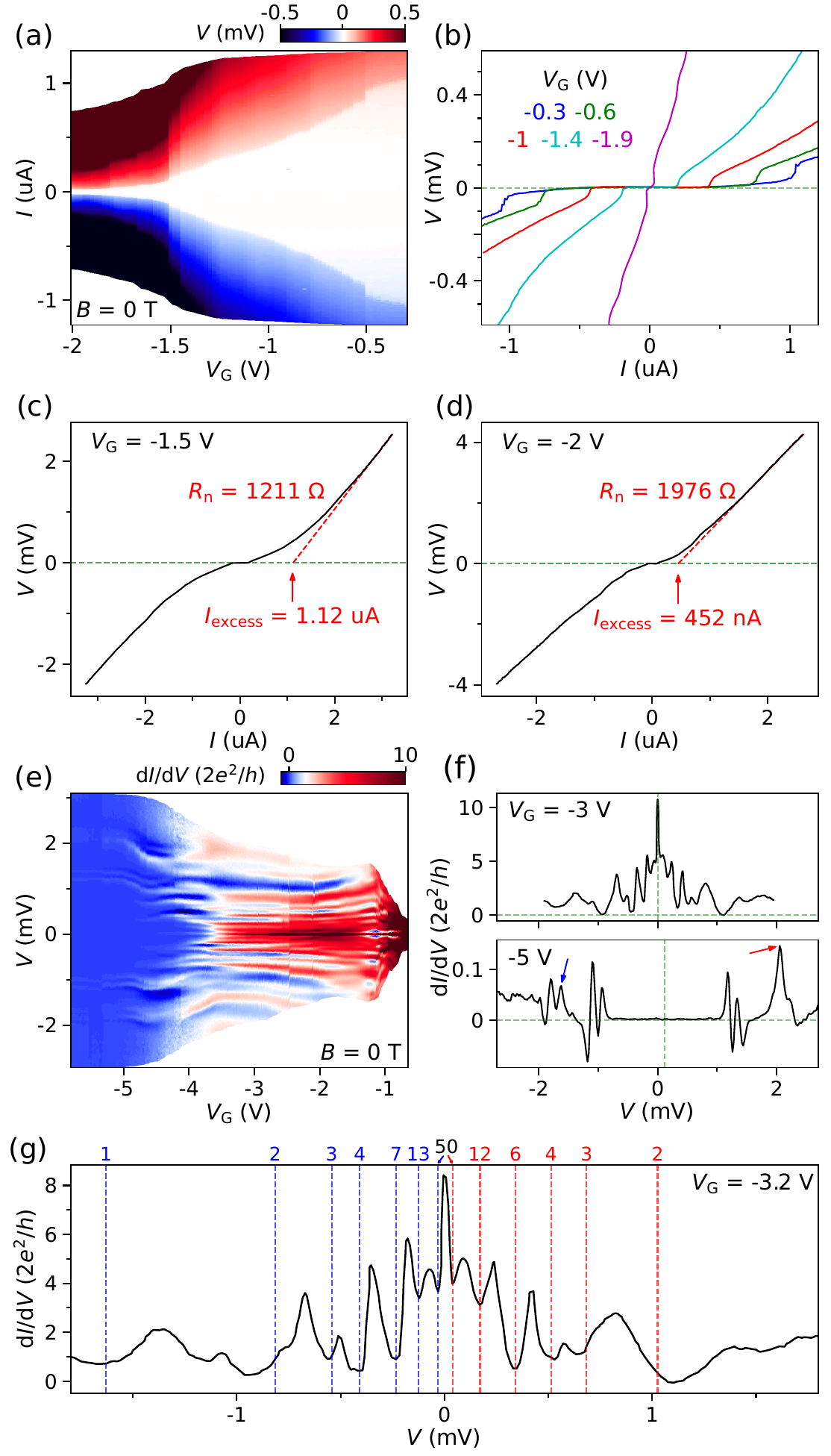}
\centering
\caption{Transport characterization of device A. (a) $I$-$V$ characteristic as a function of $V_{\text{G}}$. $B$ = 0 T. (b) Line cuts from (a). (c-d) $I$-$V$ curves over a larger bias range for the extraction of excess current. (e) $dI/dV$ as a function of $V$ and $V_{\text{G}}$. $B$ = 0 T. (f) Line cuts from (e). (g) A line cut from (e) with the dashed lines indicating positions of MARs.  }
\label{fig2}
\end{figure}

Figure 1 briefly illustrates the device design, growth procedure, and material analysis. We started with a CdTe/Pb$_{0.92}$Eu$_{0.08}$Te substrate \cite{Wenyu, Yichun, Yuhao} and covered it with a thin SiN dielectric layer.  A Hydrogen SilsesQuioxane (HSQ) shadow wall was fabricated prior to the Pb$_{0.92}$Eu$_{0.08}$Te growth \cite{Jiangyuying}. A square-shaped pattern with two additional bars (wire shape) was then defined by etching the SiN film in a process of electron beam lithography, as shown in Fig. 1(a). The square pattern defines the semiconductor 2DEG while the two extra bars are for source/drain contacts. The PbTe (blue) thin film was then selectively grown within the etched region, as depicted in Fig. 1(b). To ensure selectivity during the PbTe growth, the substrate was heated to $\sim$ 327 $^{\circ}$C. The estimated thickness of the PbTe film was $\sim$ 80 nm. This thickness is sufficient to reach the 2D or quasi-2D regime, as the subband spacing due to the confinement along this direction is estimated to be a few meV (estimated based on results in PbTe-Pb nanowires), significantly surpassing the relevant energy scales. Another estimation is based on PbTe nanowire experiments: a subband spacing of 4-5 meV is extracted from the diamond sizes of ballistic plateaus for a wire width of $\sim$ 300 nm \cite{Wenyu2024}.

For the superconductor growth, the chip was then cooled using a liquid nitrogen stage, and the Pb film was deposited at a tilted angle. This configuration allowed the HSQ wall to shadow a portion of the film, defining the junction region, as illustrated in Fig. 1(c). The entire chip was then capped with a CdTe layer to prevent oxidation of the Pb film, as shown in Fig. 1(d).

The device fabrication was similar to that in Ref. \cite{Zitong}. To prevent short-circuits, most of the Pb film on the substrate was etched using ion milling. Note that the Pb film on the PbTe device and the core region of the JJ, defined by the shadow growth, were kept intact. Electrode contacts and a side gate were then deposited (Fig. 1(e)) through evaporation of Ti/Au. Prior to the metal deposition, Ar plasma etching was performed to remove the capping layer for ohmic contacts.  Figure 1(f) shows the scanning electron micrograph (SEM) of a representative device. The Pb film (capped by CdTe) on PbTe is false-colored (violet) for clarity. The side gate was positioned in a U-shaped region where the Pb film has been etched away using ion milling.

To characterize the semiconductor-superconductor interface, we performed the scanning transmission electron microscopy (STEM) on a measured device (referred to as device A, see Figs. 2-3 for its transport characterization). Figure 1(g) shows the PbTe-Pb-CdTe layered structure. The PbTe-Pb and Pb-CdTe (capping) interfaces appear clean and sharply defined. These sharp interfaces are crucial for high-quality  quantum devices.  We further show an overview of the cross-section of the JJ in Fig. 1(h), corresponding to a cut through the midpoint of the junction, along the $y$-axis (parallel to the current flow, as depicted in Fig. 1(f)). Fig. 1(g) is an enlarged view of the red box in Fig. 1(h). The Pb film appears flat and uniform on the device scale. The middle region in Fig. 1(h) is the shadowed region (the junction), corresponding to the PbTe part not covered by Pb. The shadowed region is flat and uniformly capped by a CdTe layer, suggesting a low level of disorder. The dark and white layers above the CdTe capping are evaporated Ti/Au, not part of the device but were introduced after the measurement for the processes of focused ion beam and STEM to enhance surface conductivity. Further details regarding the material analysis (e.g. atomic resolution STEM) are provided in Fig. S1 in the Supporting Information.

Next we show the basic transport properties of device A in Fig. 2, with its corresponding SEM provided in Fig. S1. All measurements were performed in a dilution fridge at a base temperature below 50 mK. Figure 2(a) illustrates the current-voltage ($I$-$V$) characteristic and its gate ($V_{\text{G}}$) dependence at zero magnetic field ($B$ = 0 T). $I$ represents the current passing through the JJ, and $V$ denotes the voltage drop across the JJ. The white region corresponds to the supercurrent regime characterized by zero resistance, see Fig. 2(b) for line cuts. The sweeping direction of $I$ ranged from negative to positive. The measurement circuit was standard two-terminal. Thus a series resistance, contributed by the fridge filters and the device contacts, was subtracted, see Fig. S2 in the Supporting Information for a detailed analysis. The switching current $I_{\text{s}}$ can be gate-tuned from 0 to 1 $\upmu$A with $V_{\text{G}}$ varying from -2 V to -0.3 V. The retrapping current $I_{\text{r}}$, defined by the switching point in the negative bias branch, is close to $I_{\text{s}}$, suggesting an over-damped nature for device A. 

To extract the excess current ($I_{\text{excess}}$), a higher current bias was applied, as shown in Fig. 2(c). From the linear fit at $V > 2\Delta/e$, we can extract the normal state resistance $R_{\text{n}} \sim$ 1.2 k$\Omega$. The intercept of the linear fit on the $I$-axis gives the estimated $I_{\text{excess}}$, which is approximately 1.12 $\upmu$A. The size of the gap ($\Delta$) is not accurately known. If we assumed $\Delta \sim$ 1 meV, a value extracted from nanowire devices grown under identical conditions \cite{Yichun},  $eI_{\text{excess}}R_{\text{n}}/\Delta$ can be estimated to be $\sim$ 1.344, corresponding to a junction transparency of $\sim$  0.81 \cite{BTK, Flensberg_1988}. Note that this is a rough estimation due to the uncertainty of $\Delta$: if $\Delta$ of 0.815 meV was assumed (see Fig. 2(g)), the transparency would be $\sim$ 0.9. The transparency is gate-dependent, see Fig. 2(d) for another example with a transparency of $\sim$ 0.7.

\begin{figure*}[ht]
\includegraphics[width=\textwidth]{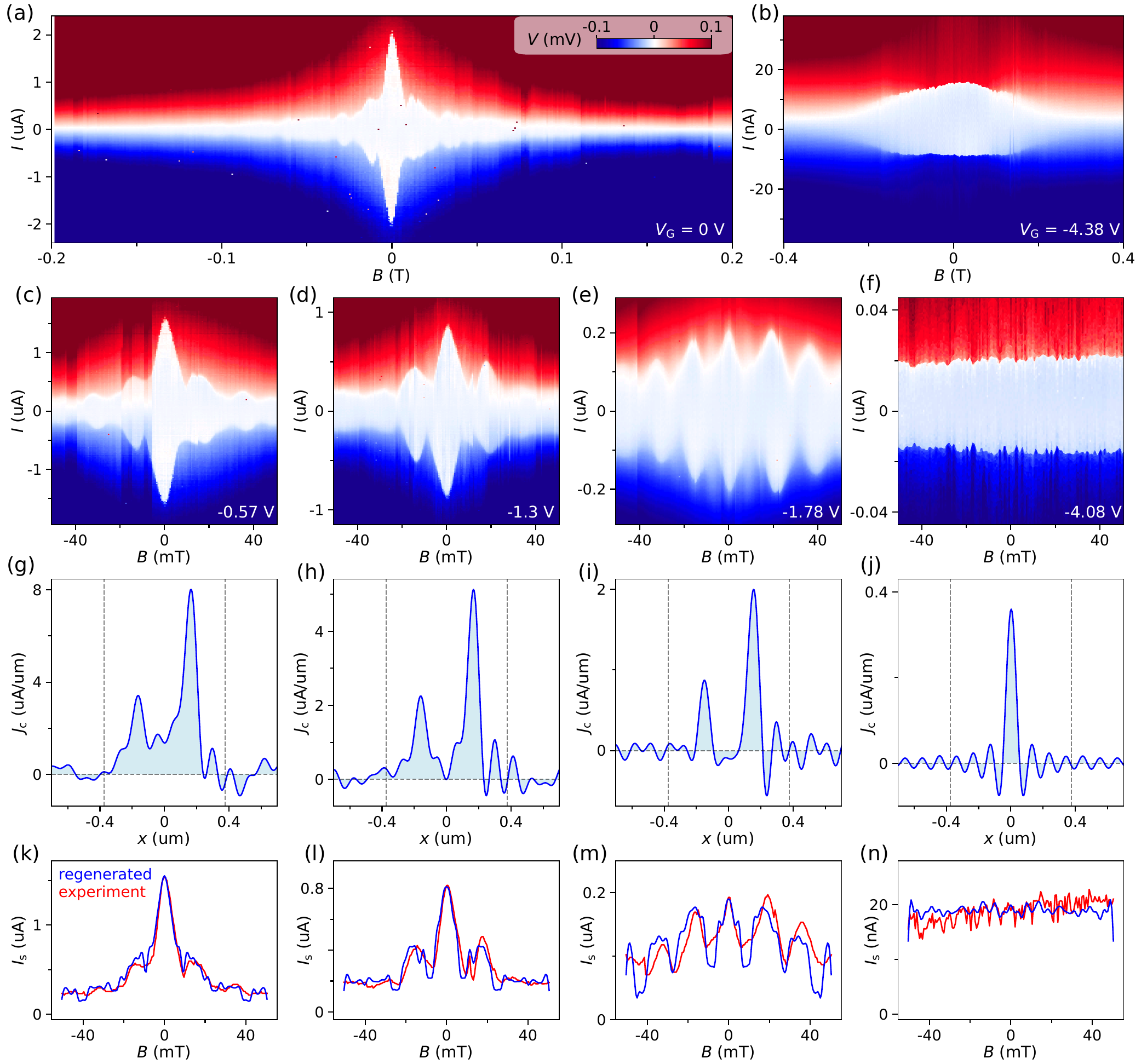}
\centering
\caption{Supercurrent interference. (a) $I$ vs $V$ and $B$ of device A. $V_{\text{G}}$ = 0 V. $B$ is perpendicular to the substrate plane unless specified. (b-f) $V_{\text{G}}$ = -4.38 V, -0.57 V, -1.3 V, -1.78 V and -4.08 V, respectively. A small offset of $B$ has been corrected to position the midpoint of the central lobe to 0 mT.  (g-j) Calculated corresponding spatial distribution of critical supercurrent density. (k-n) The red curves are $I_{\text{s}}$'s extracted from (c-f). The blue curves are the reconstructed $I_{\text{s}}$'s calculated based on the distributions in (g-j).   }
\label{fig3}
\end{figure*}

Figure 2(e) presents the differential conductance $dI/dV$ as a function of $V$ and $V_{\text{G}}$ at zero magnetic field. In the open regime (upper panel of Fig. 2(f)), a zero-bias peak is observed and attributed to supercurrent. The subgap peaks/dips are likely a result of multiple Andreev reflections (MARs) \cite{Flensberg_1988, PRApplied_MAR}. The lower panel of Fig. 2(f) corresponds to the tunneling regime. The subgap conductance is suppressed and negative differential conductance is observed near the gap edges, reminiscent of gap spectroscopy of S-NW-S devices (S for superconductor and NW for nanowire) \cite{Zitong}. However, the complexity arises here as the JJ cannot be simplified as a tunnel barrier due to its large size. The side gate also does not uniformly deplete the JJ: the junction region close to the gate gets depleted first. Additionally, the JJ shadowing is not uniform due to parasitic growth on the shadow wall (see Fig. S1 for its SEM). This inhomogeneity  can lead to a  spatially non-uniform induced gap. The line shape in the lower panel of Fig. 2(f) is not centered near zero bias,  but around $V$ = 0.12 mV (the green dashed line), possibly due to the aforementioned mechanisms. Figure 2(g) shows a line cut with dashed lines indicating the positions of MARs. Note that conductance dips (resistance peaks) should mark the positions of MARs for a JJ in the high transparency regime \cite{PRApplied_MAR}. Due to the non-zero offset of the midpoint, we assumed the value of $2\Delta$ to be 1.63 meV for the negative bias branch and 2.05 meV for the positive bias branch, corresponding to the positions of the two peaks in Fig. 2(f) (the blue and red arrows), respectively. The expected MAR positions, $2\Delta/eN$ ($N$ is an integer, see labeling), roughly align with the conductance dips with minor deviations.

We then fix the gate voltage and scan a magnetic field ($B$) that is perpendicular to the device substrate. Figure 3(a) shows such an example for device A at $V_{\text{G}}$ = 0 V. The zero-field switching current is $\sim$ 2 $\upmu$A. This value is significantly suppressed for $B$ at 0.2 T. The suppression is not monotonic, but accompanied by oscillations of $I_{\text{s}}$. These oscillations arise from the interference of the supercurrent distributed within the junction. The perpendicular $B$ induces a spatial variation of the phase gradient across the JJ along the $x$-axis (see Fig. 1(f) for its labeling), in the form of $2\pi lBx/\Phi_0$. $\Phi_0$ is the flux quantum ($h/2e$), and $l$ is the effective junction length (along the $y$-axis). The switching current of the JJ is a phase-sensitive summation (integration) of the supercurrent distributed over the $x$-axis: $I_{\text{s}}(B)=|\int_{-\infty}^{\infty} dx J_{\text{c}}(x) \text{exp}(i\frac{2\pi l B}{\Phi_0}x)|$, leading to oscillations of $I_{\text{s}}$ in $B$. $J_{\text{c}}(x)$ is the critical supercurrent density, which is a function $x$. This model assumes a sinusoidal current-phase relation. For more negative $V_{\text{G}}$ (Fig. 3(b)), the JJ exhibits a switching current of $\sim$ 20 nA, which is two orders of magnitude smaller than that of Fig. 3(a). The supercurrent is monotonically suppressed without the oscillations. This observation suggests that near the pinched-off regime, the supercurrent is ``squeezed'' into a narrow channel, where spatial variation of the phase gradient is effectively suppressed.

To elucidate the gradual evolution, we present four successive scans in Figs. 3(c-f), setting $V_{\text{G}}$ between Figs. 3(a) and 3(b) (additional scans can be found in Fig. S3 in the Supporting Information). In Fig. 3(c), the central lobe is much ``taller'' (in $I_{\text{s}}$) than the side lobes, indicating a Fraunhofer-like interference pattern where the supercurrent is distributed throughout the junction. Note that if $J_{\text{c}}$ along the $x$-axis is uniform, the width of the central lobe should be twice that of the side lobes. However, in Fig. 3(c), the central lobe width ($\sim$ 18 mT) is only slightly larger than that of the side lobes ($\sim$ 14.8 mT). This narrowing of the central lobe suggests that $J_{\text{c}}$ is not uniform along the $x$-axis. Figure 3(g) shows the calculated $J_{\text{c}}$ for Fig. 3(c), employing a method based on Fourier transform \cite{Yacoby_Fraunhofer}.  The distribution is indeed inhomogeneous and reveals two dominating peaks. The JJ edges are denoted as the two vertical dashed lines. The width of the JJ is $w \sim$ 780 nm, and we define $x$ = 0 as the JJ midpoint. We have also implemented a slightly different method from Dynes and Fulton \cite{Dynes_Fulton}, yielding a qualitatively similar distribution (see Fig. S4 in the Supporting Information).   Note that the positive direction of $x$-axis in Figs. 3(g-j) is not well-defined, i.e. it may be the same with the $x$-axis labeled in Fig. 1(f) or reversed, as this information was lost in the process of Fourier transform. 

As $V_{\text{G}}$ becomes more negative, the height of the central lobe decreases, and the oscillations become more SQUID-like, as shown in Fig. 3(e). The corresponding $J_{\text{c}}$ in Fig. 3(i) shows two peaks with other regions being depleted, confirming its ``double-slit'' nature. Based on $\Phi_0 = \Delta B \times A$ and the oscillation period $\Delta B$ = 14.8 mT, we can estimate the effective junction area $A \sim$ 0.14 $\upmu$m$^2$. In Fig. 3(f), $I_{\text{s}}$ is quite small ($\sim$ 20 nA) at more negative $V_{\text{G}}$ and the interference pattern is gone. Correspondingly, the supercurrent distribution (Fig. 3(j)) transforms into a single narrow peak. 

The origin of the two-peak feature in $J_{\text{c}}(x)$ is currently unknown. More analysis on the junction shape suggests that this feature has no correlation with the junction inhomogeneity, see Fig. S5. This is also confirmed by a second device with a more uniform junction shape. Future studies could vary the junction size and thickness to gain more insights on the underlying mechanism.

\begin{figure}[hb]
\includegraphics[width=0.8\columnwidth]{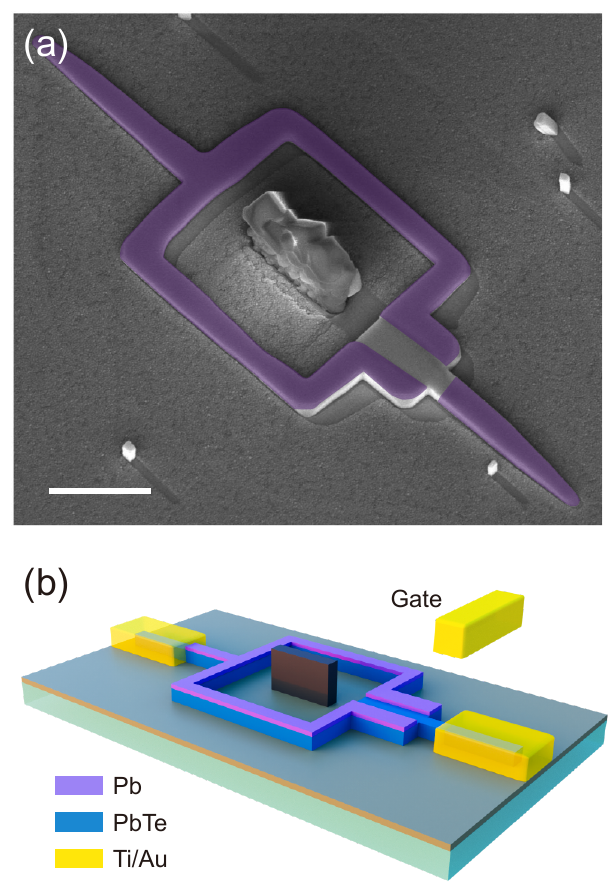}
\centering
\caption{Prototype device for Majorana detection. (a) Tilted SEM of a prototype device. The violet color highlights the Pb film on PbTe (the CdTe capping is not shown for clarity). (b) Schematic of the device with contacts and gates. The gate (top) is on a dielectric layer (not shown). }
\label{fig4}
\end{figure}

In Figs. 3(k-n), we reconstruct the oscillation patterns of $I_{\text{s}}$ using the calculated $J_{\text{c}}(x)$ in Figs. 3(g-j) as an input. Plugging this $J_{\text{c}}$ into the formula, we reconstruct $I_{\text{s}} (B)$ (the blue curves), and compare it to the $I_{\text{s}}$ (the red curves) extracted from Figs. 3(c-f). The qualitative agreement suggests that the calculated $J_{\text{c}}$ is reliable. Note that we did not consider the London penetration depth and the effect of flux focusing, i.e. $l$ is purely the spacing between the two Pb films. Including those effects would cause a smaller spacing of the two peaks in $J_{\text{c}}$ in Figs. 3(g-i). Figs. S6 and S7 show three additional devices exhibiting gate-tunable supercurrents. In Fig. S8, we show a second device, where the interference can be tuned from Fraunhofer-like to SQUID-like. The effect of the two-peak feature in the SQUID-like regime, in relation to Majorana detection, remains to be addressed by theory. Experimentally, one could tune the device into the Fraunhofer-like regime with a uniform current distribution for Majorana detection, and then gradually move to the SQUID-like regime to study this effect.

After establishing the feasibility of the PbTe-Pb hybrid as a planar JJ platform, Figure 4 presents a prototype device designed for future Majorana detection. To enable the application of phase bias, the two bars of the planar JJ are interconnected, forming a PbTe nanowire network loop, as depicted in Fig. 4(a). The shadow wall, fabricated within the loop, defines the JJ region on the PbTe film. The violet region represents the Pb film (capped by CdTe) on PbTe. The nanowire loop and the planar JJ together form an ac SQUID. The phase drop across the planar JJ can be tuned by the flux penetrating through the loop, further controlled by a small out-of-plane magnetic field. 

Figure 4(b) shows the schematic of the device. The source contact is deposited on the PbTe nanowire, connecting to one edge of the planar JJ for probing its edge state. The drain contact connects to the ac SQUID. Top gates can be fabricated after the deposition of a dielectric layer (not drawn for clarity). One top gate can be used to tune the barrier height in the PbTe nanowire region between the source contact and the edge of the JJ. Another gate (not drawn for clarity) can be deposited on top of the junction region to tune its electrochemical potential. This device setup, achieved mainly through selective area growth, is equivalent to Ref. \cite{2019_Nature_Charlie, 2019_Nature_Yacoby}, where tunneling spectroscopy can be performed to detect a single Majorana zero mode. For the detection of two Majoranas and their correlation, three-terminal devices should also be possible through a more sophisticated geometry of the shadow wall and the network loops. 
  
In summary, we have demonstrated the selective area growth of PbTe-Pb hybrid planar Josephson junctions as a potential Majorana material platform. Unlike the previous 2DEG top-down approach, the junction region was formed during the growth process instead of by wet etching. This approach allows us to  keep the core region of the JJ intact and may reduce disorder. Basic transport calibration reveals a gate-tunable supercurrent and supercurrent interference in a perpendicular magnetic field, characteristics for a planar JJ. We also present the growth of a prototype device where the phase bias can be applied and tunneling spectroscopy can be performed to detect Majorana zero modes.  

\section{Acknowledgment} 

We thank Hechen Ren for valuable comments. This work is supported by Tsinghua University Initiative Scientific Research Program, National Natural Science Foundation of China (92065206) and the Innovation Program for Quantum Science and Technology (2021ZD0302400). 
 
\section{Data Availability} 

Raw data and processing codes within this paper are available at https://doi.org/10.5281/zenodo.10903435

\section{References}

\bibliography{mybibfile}

\newpage

\onecolumngrid

\newpage


\section*{Supplementary material}

# Supporting Information for "Selective-Area-Grown PbTe-Pb Planar Josephson Junctions for Quantum Devices"

Ruidong Li,[1, *] Wenyu Song,[1, *] Wentao Miao,[1, *] Zehao Yu,[1, *] Zhaoyu Wang,[1] Shuai Yang,[1] Yichun Gao,[1]
Yuhao Wang,[1] Fangting Chen,[1] Zuhan Geng,[1] Lining Yang,[1] Jiaye Xu,[1] Xiao Feng,[1, 2, 3, 4] Tiantian Wang,[2, 4]
Yunyi Zang,[2, 4] Lin Li,[2] Runan Shang,[2, 4] Qi-Kun Xue,[1, 2, 3, 4, 5] Ke He,[1, 2, 3, 4, †] and Hao Zhang[1, 2, 3, ‡]

[1] *State Key Laboratory of Low Dimensional Quantum Physics,*
*Department of Physics, Tsinghua University, Beijing 100084, China*
[2] *Beijing Academy of Quantum Information Sciences, Beijing 100193, China*
[3] *Frontier Science Center for Quantum Information, Beijing 100084, China*
[4] *Hefei National Laboratory, Hefei 230088, China*
[5] *Southern University of Science and Technology, Shenzhen 518055, China*

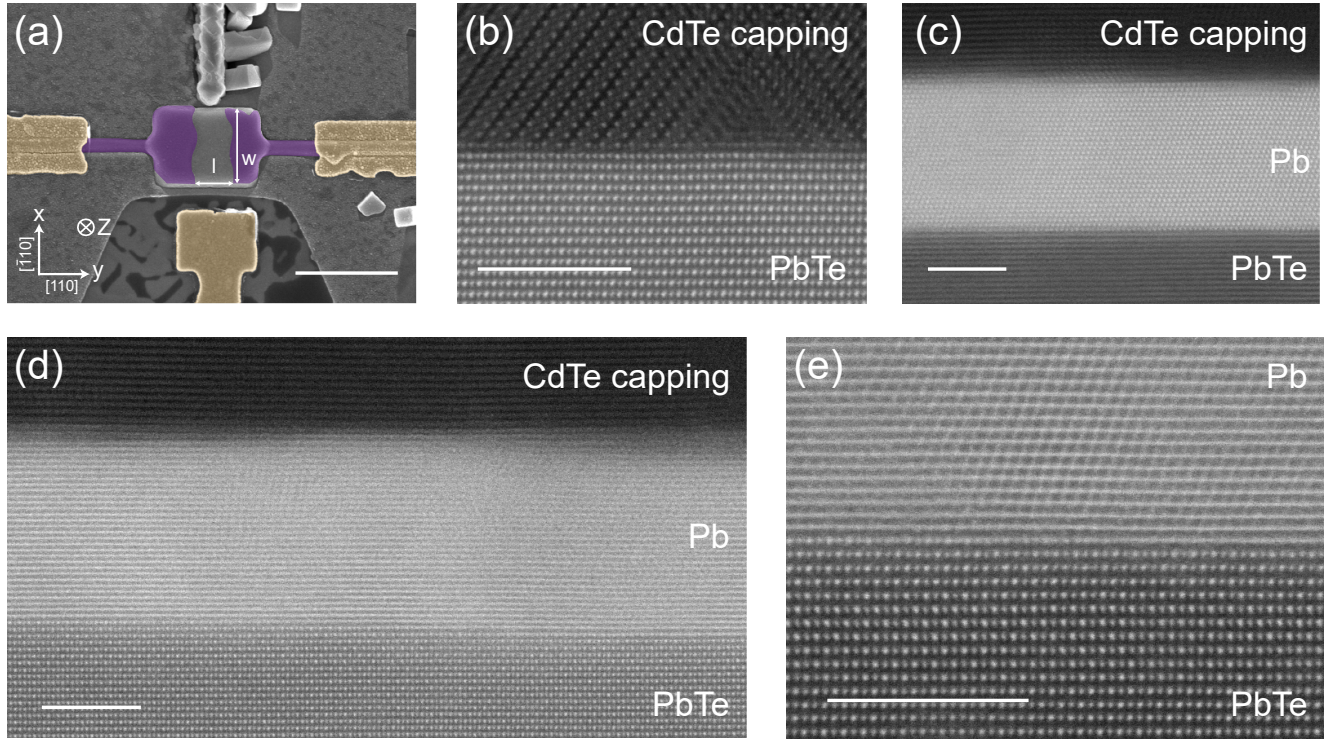

FIG. S1. (a) SEM of device A. The scale bar is 1 μm. The Pb film (capped with CdTe) on PbTe is highlighted (violet). The width ($w$) and length ($l$) of the junction are labeled, as well as the coordinate axis and the crystal orientation. (b-e) Atomic resolution TEM of device A. The scale bar is 5 nm. The measurement circuit is two-terminal, identical to that in Refs. [38–41]. Thus bias drop over the series resistance, $R_{series} = R_{fridge} + R_{contact}$, has been subtracted from $V$. $R_{fridge}$ is the fridge filter resistance and input/output resistance of the circuit, calibrated independently by shorting two fridge lines at the mixing chamber, see Fig. S2(d). We find that $R_{contact}$, the device contact resistance, varies for different $V_G$'s, possibly due to the gate-tunable PbTe regions underneath the superconductor. $R_{fridge}$ is 250 Ω for Figs. 2(a,b,e,f,g), 443, and 627 Ω for Figs. 2(c) and 2(d), respectively. For Figs. 3(a-f), it is 365, 2132, 354, 370, 380, and 1595 Ω, respectively.

---

* equal contribution
† kehe@tsinghua.edu.cn
‡ hzquantum@mail.tsinghua.edu.cn

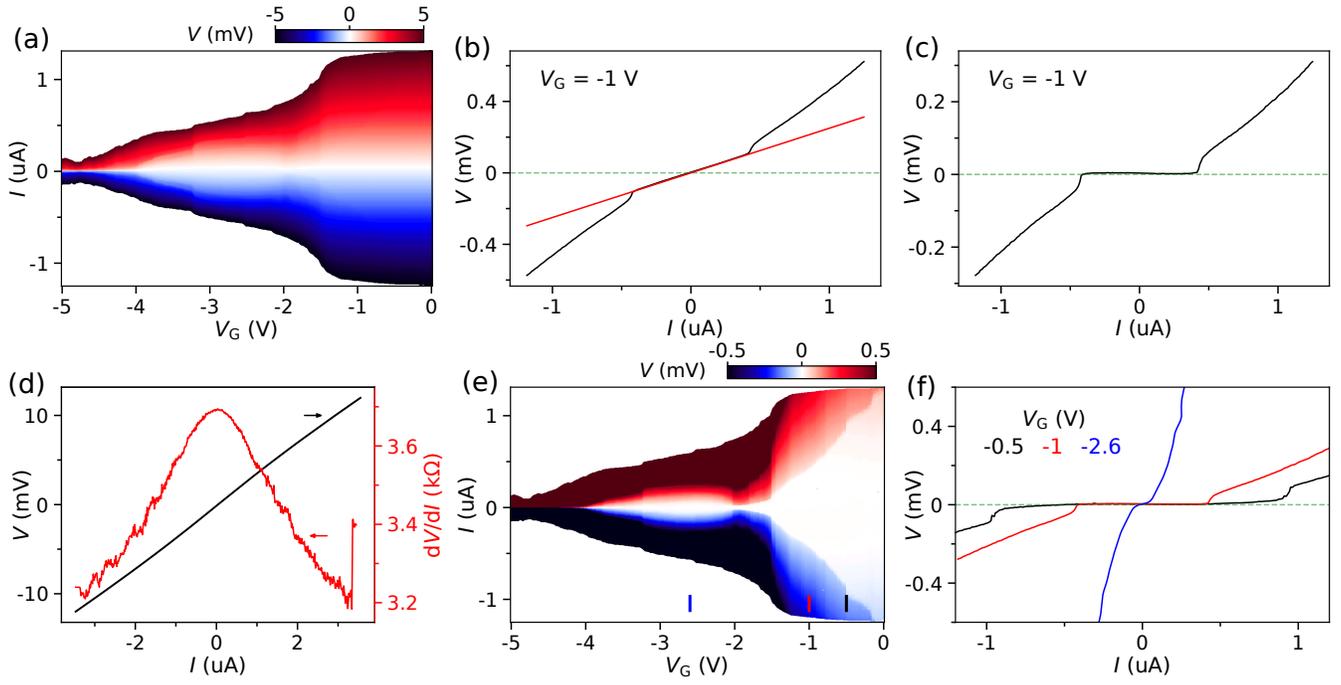

FIG. S2. (a) *I-V* of device A without subtracting $R_{\text{contact}}$. (b) A line cut (the black curve) from panel (a). The jumps around $\pm$ 0.5 $\mu$A are the switching points between superconducting and resistive states. The red curve is a linear fit of the superconducting states, corresponding to $R_{\text{contact}} \sim 250$ $\Omega$. (c) The *I-V* curve after subtracting the background contributed by $R_{\text{contact}}$. (d) *I-V* characteristic of $R_{\text{fridge}}$ by shorting two fridge lines at the mixing chamber. The data in panel (a) was obtained by subtracting this $R_{\text{fridge}}$ from the raw two-terminal data. The red curve (right axis) is the differential resistance $dV/dI$ (after a minor smoothing), obtained by numerical derivative. (e) *I-V* after subtracting $R_{\text{fridge}}$ and $R_{\text{contact}}$. Figure 2(a) corresponds to the right part of panel (e). (f) Three line cuts from (e).

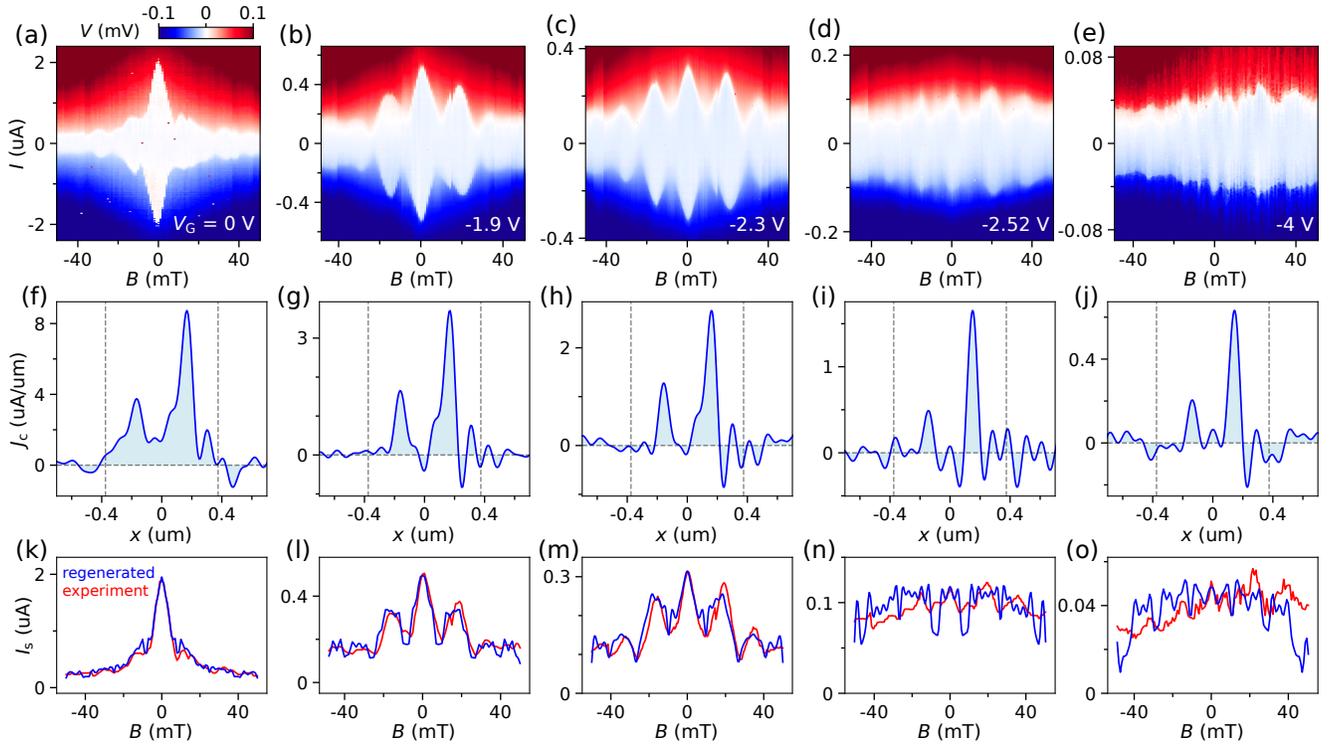

FIG. S3. Additional $B$ scans with $V_G$ labeled in each panel. The second row is the calculated critical supercurrent distribution while the third row is the regenerated oscillation pattern (blue) and the extracted $I_s$ from (a-e) (red). Panel (a) is the same data of Fig. 3(a). $R_{contact}$ is 365, 343, 389, 428, and 753 $\Omega$ for panels (a-e), respectively.

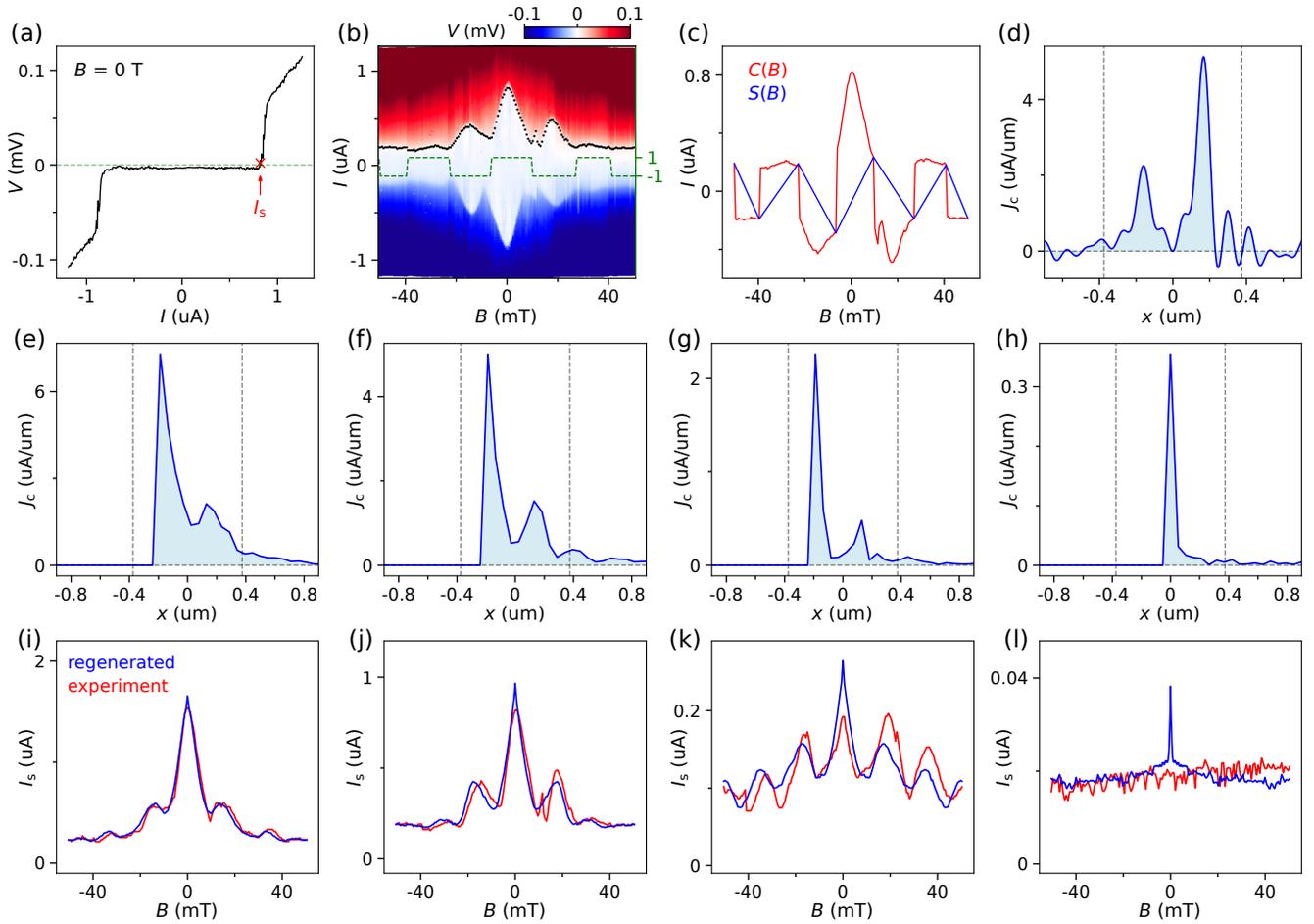

FIG. S4. Analysis details of the critical current distribution. (a) $I_s$ extracted from an $I$-$V$ curve, see the red arrow. (b) Re-plotting Fig. 3(d) with the black dots marking $I_s$. The green dashed line is a flip function that changes sign at each node of the lobe, a method following Ref. [1]. (c) Calculated even component ($C(B)$) and odd component ($S(B)$) of $I_s(B)$. $C(B)$ is obtained by multiplying the black dots in (b) with the green dashed line, while $S(B)$ is the linear interpolation between the node points of $C(B)$. We then can calculate $J_c(x = \frac{L}{\Phi_0} \int_{-\infty}^{\infty} dB [C(B) + iS(B)] \exp(-i\frac{2\pi lB}{\Phi_0} x)$ with result shown in (d). The integral is converted to a discrete summation. (e-h) Calculated $J_c(x)$ using the method of Dynes and Fulton [2, 3] for Figs. 3(c-f), where a discrete Hilbert transform is used to construct the phase. The transition from Fraunhofer-like to SQUID-like is consistent with the results in Figs. 3(g-j) where the method in Ref. [1] was used. (i-l) Calculated $I_s$ oscillations (blue) using the $J_c(x)$ in (e-h) as an input. The red curves are the extracted $I_s$ from Figs. 3(c-f). The deviation is larger than that in Figs. 3(k-n).

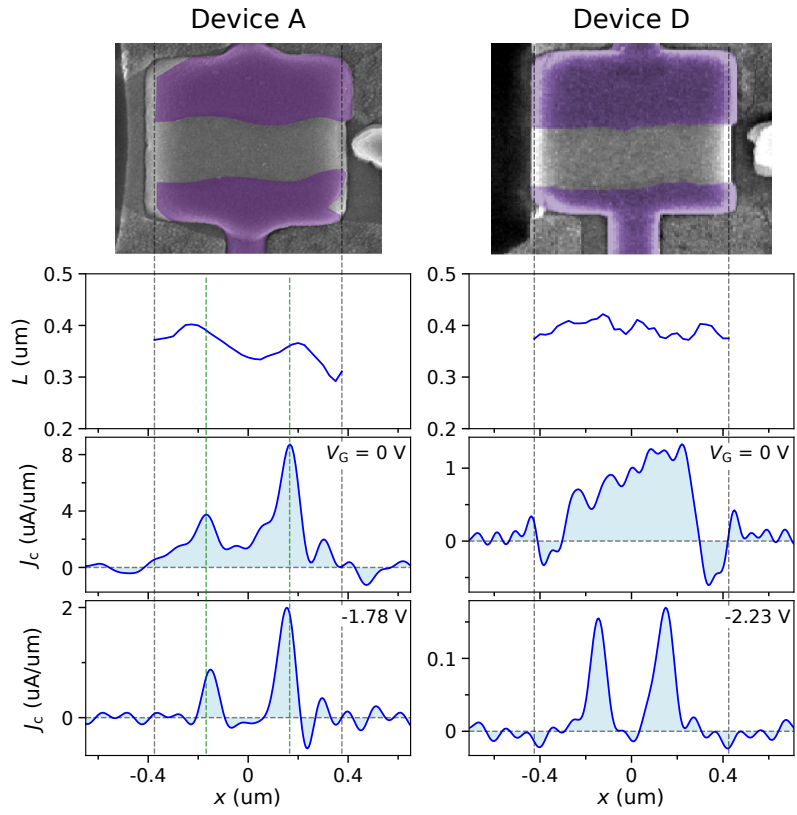

FIG. S5. Analysis on the junction inhomogeneity and current distribution. First row, enlarged SEMs of devices A and D. Second row, contact spacing (between two Pb films) as a function of $x$. The black dashed lines mark the junction edges. Third and fourth rows, extracted current distributions for different $V_G$'s. The two-peak feature (green dashed lines) of device A is not correlated with the profile variation of the contact spacing. Moreover, device D has a more uniform junction shape, and still exhibits the two-peak feature in the SQUID-like regime, suggesting its intrinsic nature.

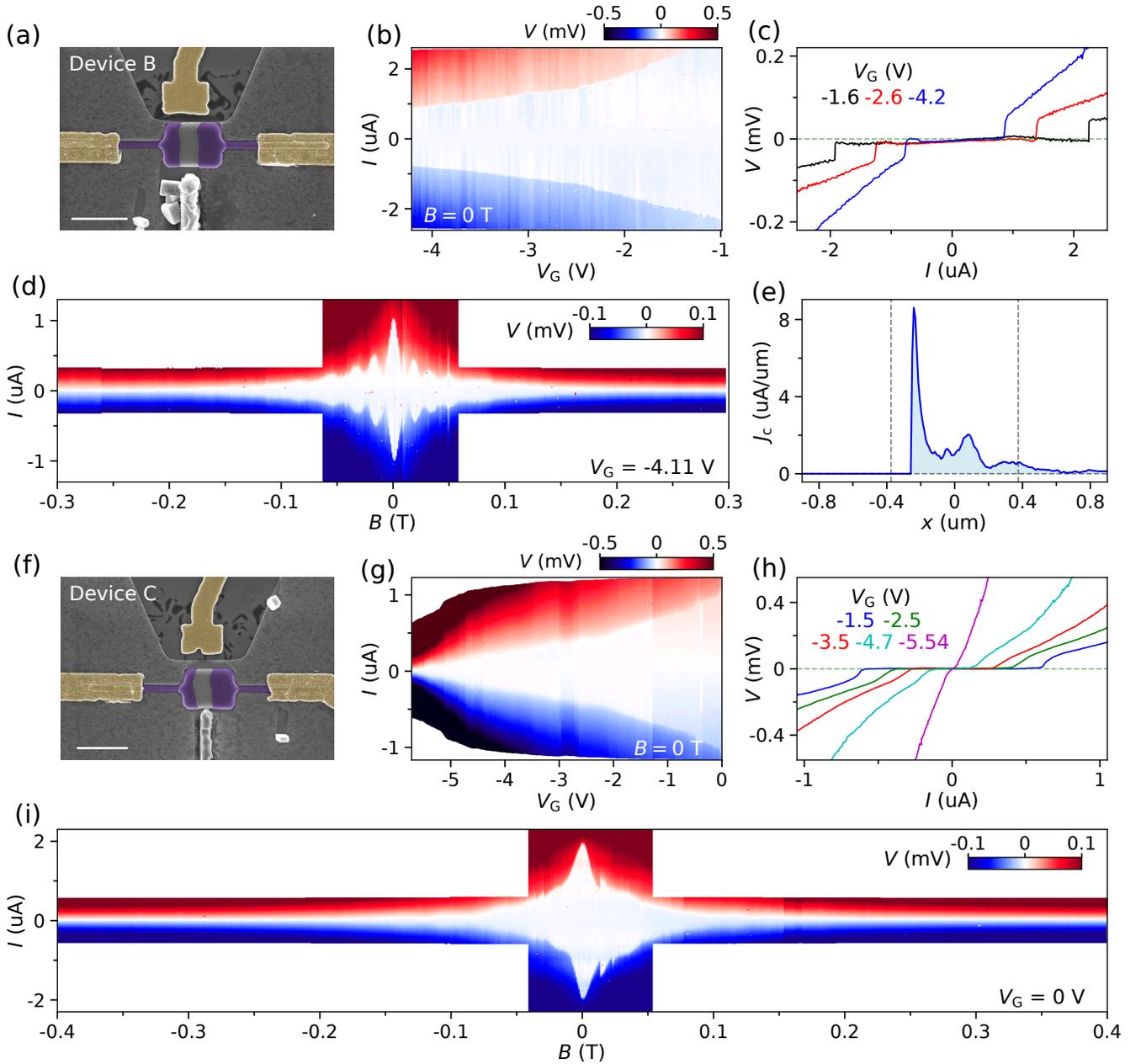

FIG. S6. (a) SEM of a second planar JJ device (device B). The scale bar is 1 μm. (b) Gate dependence of the supercurrent in device B. More negative gate voltages ($V_G < $ -4 V) was not applied due to leakage. (c) Line cuts from (b). The fluctuations at low current bias is due to the inaccuracy of the $R_{\text{fridge}}$ estimation. (d) Interference of $I_s$ as a function a perpendicular $B$ in device B. (e) Calculated $J_c(x)$ corresponding to (d). (f-i) Measurement of a third device (device C), similar to (a-d). $J_c$ cannot be calculated for (i) due to the absence of supercurrent oscillations and nodes. The scale bar is 1 μm. $R_{\text{contact}}$ is 535, 544, 376, and 375 Ω for panels (b), (d), (g) and (i), respectively.

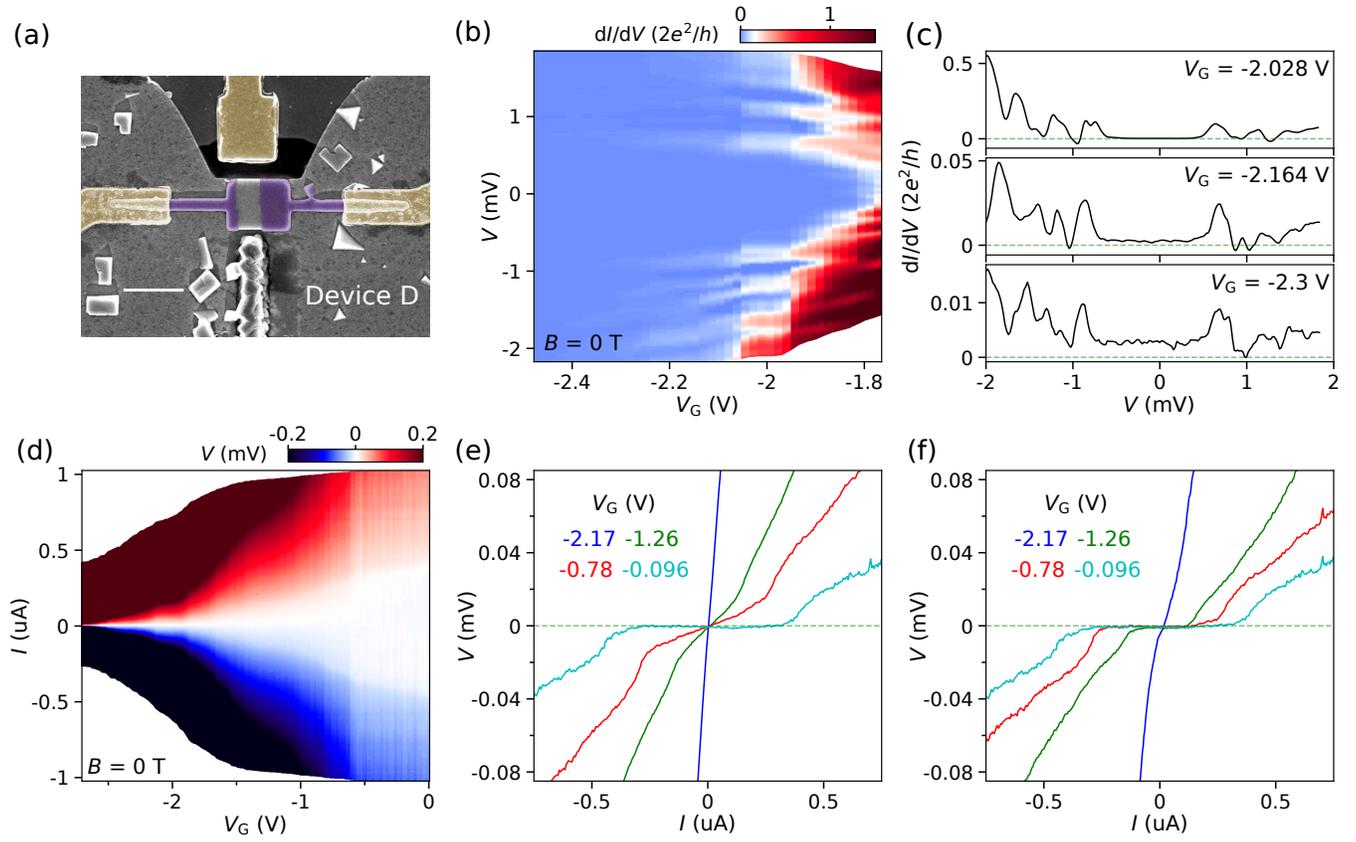

FIG. S7. False-color SEM of device D. Scale bar, 1 µm. (b) $dI/dV$ vs $V$ and $V_G$ in the tunneling regime, $B = 0$ T. (c) Line cuts from (b). (d) $V$ vs $I$ and $V_G$, illustrating the gate-tunable supercurrent (white regions). $B = 0$ T. (e) Line cuts from (d). The tilted supercurrent regions suggest a gate-tunable contact resistance $R_{\text{contact}}$. A fixed $R_{\text{contact}} = 300\ \Omega$ was assumed for (b-e). (f) Line cuts assuming a gate-dependent $R_{\text{contact}}$.

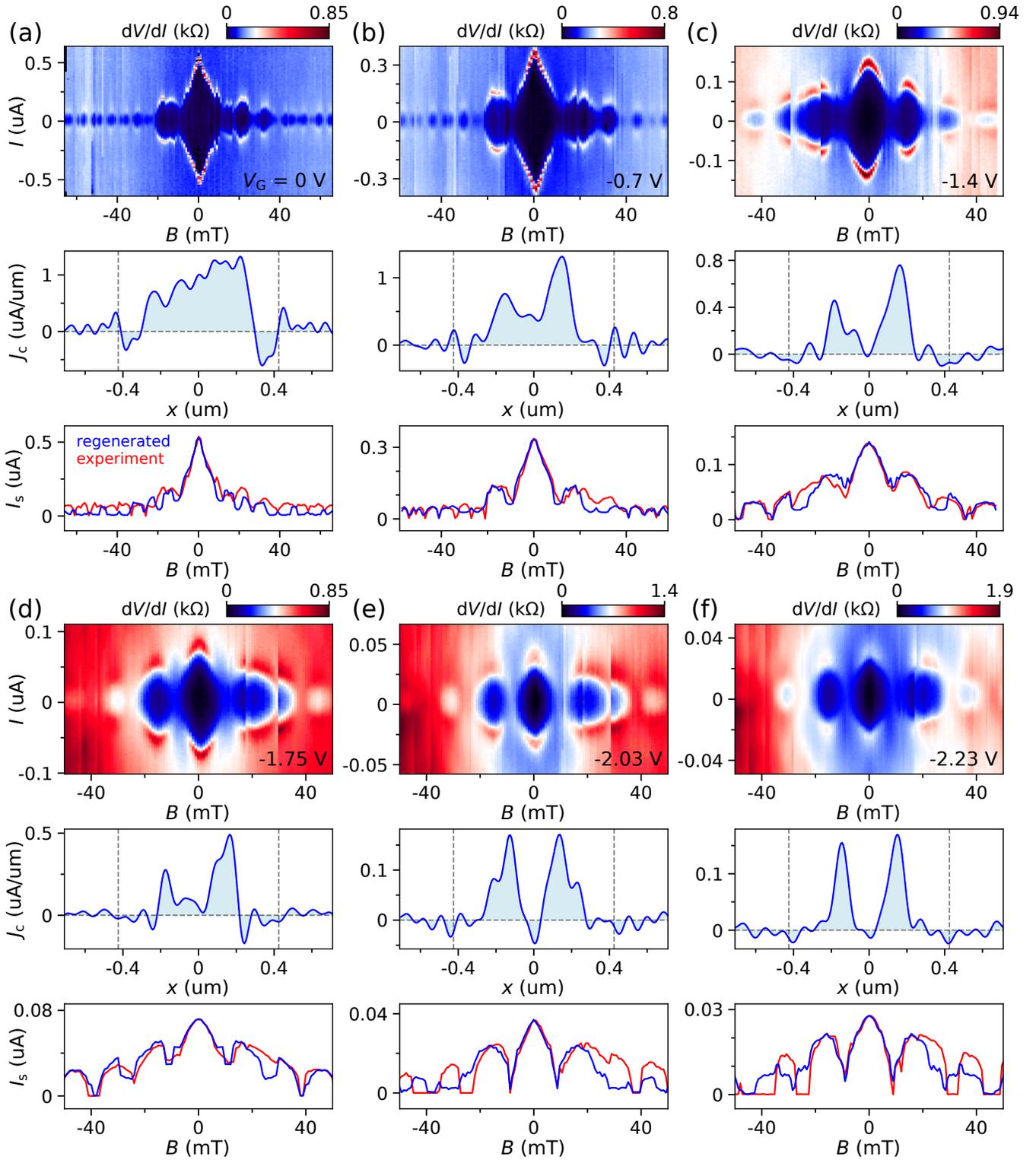

FIG. S8. Supercurrent interference in device D. (a-f) are results at 6 fixed $V_G$'s (see labeling). Upper panels are the interference patterns. Middle panels are the corresponding $J_c(x)$ after the Fourier transform. Lower panels are the $I_s$ (red) extracted from the upper panels, and the comparison to the regenerated $I_s$ (blue) based on the middle panels. As $V_G$ decreases, the interference pattern evolves from Fraunhofer-like to SQUID-like, also evident in $J_c(x)$.



\end{document}